\documentclass[12pt]{article}
\newcommand{\nc}{\newcommand}
\nc{\bib}{\bibitem}
\nc{\al}{\alpha}
\nc{\g}{\gamma}
\nc{\G}{\Gamma}
\nc{\D}{\Delta}
\nc{\eps}{\epsilon}
\nc{\la}{\lambda}
\nc{\La}{\Lambda}
\nc{\var}{\varphi}
\nc{\cg}{{\cal G}}
\nc{\pa}{\partial}
\nc{\nn}{\nonumber \\ }
\nc{\hf}{\frac{1}{2}}  
\nc{\dz}{\frac{dz}{2\pi i}}
\nc{\bin}[2]{\left (\begin{array}{c} {#1}\\ {#2} \end{array}\right )}
\nc{\ben}{\begin{equation}}
\nc{\een}{\end{equation}}
\nc{\bea}{\begin{eqnarray}}
\nc{\eea}{\end{eqnarray}}
\nc{\bra}[1]{\langle {#1}|}
\nc{\ket}[1]{|{#1}\rangle}
\newcommand{\Z}{\mbox{$Z\hspace{-2mm}Z$}}
\nc{\C}{\mbox{\hspace{1.24mm}\rule{0.2mm}{2.5mm}\hspace{-2.7mm} C}}
\nc{\Nat}{\mbox{\hspace{.04mm}\rule{0.2mm}{2.8mm}\hspace{-1.5mm} N}}
\nc{\HH}{\mbox{\hspace{.04mm}\rule{0.2mm}{2.8mm}\hspace{-1.5mm} H}}

\def\vvdots{\mathinner{\mkern1mu\raise1pt\vbox{\kern7pt\hbox{.}}\mkern2mu
 \raise4pt\hbox{.}\mkern2mu\raise7pt\hbox{.}\mkern1mu}}

\begin{document}

\topmargin -5mm
\oddsidemargin 5mm

\begin{titlepage}
\setcounter{page}{0}

\vspace{8mm}
\begin{center}
{\huge Stochastic evolutions in superspace}\\[.4cm]
{\huge and superconformal field theory}

\vspace{15mm}
{\Large J{\o}rgen Rasmussen}\\[.3cm] 
{\em Centre de recherches math\'ematiques, Universit\'e de Montr\'eal}\\ 
{\em Case postale 6128, 
succursale centre-ville, Montr\'eal, Qc, Canada H3C 3J7}\\[.3cm]
rasmusse@crm.umontreal.ca

\end{center}

\vspace{10mm}
\centerline{{\bf{Abstract}}}
\vskip.4cm
\noindent
Some stochastic evolutions of conformal maps can be described by
SLE and may be linked to conformal field theory via stochastic
differential equations and singular vectors in highest-weight modules
of the Virasoro algebra. Here we discuss how this may be extended
to superconformal maps of $N=1$ superspace with links to superconformal
field theory and singular vectors of the $N=1$ superconformal
algebra in the Neveu-Schwarz sector. 
\\[.5cm]
{\bf Keywords:} Stochastic evolutions, 
 superconformal field theory, superspace. 
\end{titlepage}
\newpage
\renewcommand{\thefootnote}{\arabic{footnote}}
\setcounter{footnote}{0}

\section{Introduction}

Stochastic L\"owner evolutions (SLEs) have been introduced by Schramm
\cite{Sch} and further developed in \cite{LSW,RS} as a mathematically
rigorous way of describing certain two-dimensional systems
at criticality.
The method involves the study of stochastic evolutions of conformal maps.

An intriguing link to conformal field theory (CFT) has been examined
by Bauer and Bernard \cite{BB} (see also \cite{FW}) in which the SLE
differential equation is associated to a particular random walk on the 
Virasoro group. The relationship can be made more direct by establishing
a connection between the representation theory of CFT
and quantities conserved in mean under the stochastic
process. This is based on the existence of level-two
singular vectors in the associated Virasoro highest-weight modules.

In \cite{LR} it is discussed how SLE appears naturally at grade
one in a hierarchy of SLE-type growth processes.
Extending the approach of Bauer and Bernard, each of these may be
linked to CFT, and a direct relationship to the Yang-Lee singularity
is found at grade two through
the construction of a level-four null vector.

The objective of the present work is to extend this to
a link between stochastic evolutions of superconformal maps in $N=1$ 
superspace and $N=1$ superconformal field theory (SCFT).
It can be made more direct by relating some quantities conserved in
mean under the process to a level-3/2 singular
vector in the Neveu-Schwarz sector of the SCFT.
These results rely on an Ito calculus, not just for
commuting and anti-commuting variables as discussed in \cite{Rog}
and references therein, but for general non-commuting objects (e.g.
group- or algebra-valued entities) on superspace.
The generalization is straightforward, though.

After a brief review on SLE and the approach of Bauer and Bernard, 
we summarize some basics 
on $N=1$ superspace and $N=1$ SCFT in the superfield formalism.
We then discuss how graded stochastic evolutions may be
linked to SCFT and establish relationships
through two different ways of obtaining level-3/2
singular vectors. Some coupled stochastic differential
equations are solved in the process. In one case, the solution
suggests the introduction of an SLE-type trace and may be
interpreted as describing the evolution of hulls in the complex
plane. The final section contains
some concluding remarks.

\section{SLE and CFT}

Geometrically, (chordal) SLE \cite{Sch,LSW,RS} describes
the evolution of boundaries of simply-connected domains
in the complex plane. 
One may think of this as Brownian motion on the set of 
conformal maps $\{g_t\}$ satisfying the L\"owner equation
\ben
 \pa_tg_t(z)\ =\ \frac{2}{g_t(z)-\sqrt{\kappa}B_t}\ ,\ \ \ \ 
  \ \ \ \ \ g_0(z)\ =\ z
\label{g}
\een
$B_t$ is one-dimensional Brownian motion with $B_0=0$ and
$\kappa\geq0$. This is the only adjustable parameter
thus characterizing the process which is often denoted SLE$_\kappa$.

Introducing the function
\ben
 f_t(z):=\ g_t(z)-\sqrt{\kappa}B_t
\label{f}
\een
one has the stochastic differential equation
\ben
 df_t(z)\ =\ \frac{2}{f_t(z)}dt-\sqrt{\kappa}dB_t\ ,\ \ \ \ 
    \ \ \ \ \ f_0(z)\ =\ z
\label{df}
\een
based on which we may establish contact with CFT.
One defines the SLE trace as
\ben
 \g(t):=\ \lim_{z\rightarrow0}f^{-1}(z)
\label{trace}
\een
As discussed in \cite{RS}, its nature depends radically
on $\kappa$.

To unravel the link \cite{BB} one considers
Ito differentials of Virasoro group elements $\cg_t$:
\ben
 \cg_t^{-1}d\cg_t\ =\ \left(-2L_{-2}+\frac{\kappa}{2}L_{-1}^2\right)dt
  +\sqrt{\kappa}L_{-1}dB_t\ ,\ \ \ \ \ \ \ \cg_0=1
\label{G}
\een
Such group elements are obtained by exponentiating
generators of the Virasoro algebra 
\ben
 [L_n,L_m]\ =\ (n-m)L_{n+m}+ \frac{c}{12}n (n^2-1) \delta_{n+m,0}
\label{vir}
\een
where $c$ is the central charge.
The conformal transformation generated by a Virasoro group element
has a simple action on a primary field of weight $\Delta$.
In particular, the transformation generated by $\cg_t$  
reads\footnote{For simplicity,
we do not distinguish explicitly between boundary and bulk primary
fields, nor do we write the anti-holomorphic part.}
\ben
 \cg_t^{-1}\phi_\Delta (z)\cg_t\ 
  =\ \left(\partial_z f_t(z)\right)^\Delta \phi_\Delta (f_t(z))
\label{GphiG}
\een
for some conformal map $f_t$.
Using that the Virasoro generators act as
\ben
 [L_n,\phi_\Delta(z)]\ =\ (z^{n+1}\partial_z + \Delta(n+1)z^n) \phi_\Delta(z)
\label{Lphi}
\een
one finds that the conformal map $f_t$ associated to the random process 
$\cg_t$ (\ref{G}) must be a solution to the stochastic differential equation
(\ref{df}).

This link does not teach us to which CFT with central charge
$c$ a given SLE$_\kappa$ may be associated. 
A refinement could be reached, though, by determining
for which $\kappa$ the stochastic process (\ref{G}) would allow
us to relate conformal correlation functions on one
side to probabilities of stopping-time events on the other side.
A goal is thus to find
the evolution of the expectation values of the observables
of the stochastic process (\ref{G}) where the observables
are thought of as functions on the Virasoro group.
The direct link is established by relating the representation theory of
the Virasoro algebra to quantities conserved in mean under
the random process.

To achieve this, let $\ket{\D}$ denote the highest-weight 
vector of weight $\Delta$ in the Verma module ${\cal V}_\Delta$, 
and consider the time evolution of the
expectation value of $\cg_t\ket{\D}$:
\ben
 \pa_t{\bf E}[\cg_t\ket{\Delta}]\ =\ {\bf E}[\cg_t\left(-2L_{-2}
  +\frac{\kappa}{2}L_{-1}^2\right)\ket{\Delta}]
\label{LD}
\een
For some values of $\kappa$ (in relation to the central charge $c$),
the linear combination $-2L_{-2}+\frac{\kappa}{2}L_{-1}^2$ will
produce a singular vector when acting on the highest-weight vector. 
This happens provided 
\ben
 c_\kappa\ =\ 1-\frac{3(4-\kappa)^2}{2\kappa}\ ,\ \ \ \ \ \ 
 \Delta_\kappa\ =\ \frac{6-\kappa}{2\kappa}
\label{cd}
\een
In this case the expectation value (\ref{LD}) vanishes
in the module obtained by factoring out
the singular vector from the reducible Verma module ${\cal V}_\Delta$.
The representation theory of the factor module is
thereby linked to the description of quantities conserved
in mean.

We see that this direct relationship between SLE$_\kappa$ 
and CFT is through the existence of a level-two
singular vector in a highest-weight module.
This has been extended in \cite{LR} where
a link between elements of 
a hierarchy of SLE-type growth processes 
and CFT has been established. At grade two, a direct relationship
between an SLE-type stochastic differential equation 
and the Yang-Lee singularity is based on a
level-four null vector. Ordinary SLE appears
at grade one in this hierarchy.

In the following we shall discuss how 
the scenario outlined above may be extended
to stochastic evolutions in superspace with links to SCFT.

\section{Superspace and SCFT}

Here we shall give a very brief summary of results 
from the theory of
$N=1$ superspace and $N=1$ SCFT required in
the following.
We refer to \cite{Dor,Nag} and references therein for 
recent accounts on the subject.
As we only consider $N=1$ superspace and $N=1$ SCFT
we shall simply refer to them as superspace and SCFT,
respectively.

\subsection{Superconformal maps}

Let 
\ben
 (z,\theta)\ \mapsto\ (z',\theta') \ ,\ \ \ \ \ \ \ \ \ \left\{
   \begin{array}{ll} z'\ =\ g(z)+\theta\g(z)\\ \mbox{}\\
       \theta'\ =\ \tau(z)+\theta s(z) \end{array} \right.
\label{zt}
\een
denote a general superspace coordinate transformation. 
The associated superderivative reads
\ben
 D\ =\ \pa_\theta+\theta\pa_z
\label{D}
\een
Here $\theta$, $\theta'$, $\g$ and $\tau$ are anti-commuting
or (Grassmann) odd entities, while $z$, $z'$, $g$ and $s$ are even.
A superconformal transformation may be characterized
by
\ben
 Dz'\ =\ \theta' D\theta'
\label{Dz'}
\een
in which case we have
\ben
 \g(z)\ =\ \tau(z)s(z)\ ,\ \ \ \ \ \ \pa_z g(z)\ =\ s^2(z)-\tau(z)\pa_z\tau(z)
\label{gts}
\een
and 
\ben
 D\ =\ (D\theta')D'\ ,\ \ \ \ \ \ D'\ =\ \pa_{\theta'}+\theta'\pa_{z'}
\label{D'}
\een
We shall be interested in locally invertible superconformal
maps implying that the complex part of $z'$ is non-vanishing
if the complex part of $z$ is non-vanishing.
We assume this is the case for $z\neq0$.

The superconformal transformations are generated
by $T$ and $G$ with modes $L_n$ and $G_r$
satisfying the superconformal algebra
\bea
 \left[L_n,L_m\right]&=&(n-m)L_{n+m}+ \frac{c}{12}n (n^2-1) \delta_{n+m,0}\nn
 \left[L_n,G_r\right]&=&(\frac{n}{2}-r)G_{n+r}\nn
 \left\{G_r,G_s\right\}&=&2L_{r+s}+\frac{c}{3}(r^2-\frac{1}{4})\delta_{r+s,0}
\label{LG}
\eea
The algebra is said to be in the Neveu-Schwarz (NS) sector when
the supercurrent $G$ has half-integer modes, and in the
Ramond sector when they are integer. We shall consider
the NS sector only.

\subsection{Primary superfields and singular vectors}

A superfield depends on the superspace coordinates
and thus enjoys an almost trivial Taylor expansion
in $\theta$:
\ben
 \Phi(z,\theta)\ =\ \phi(z)+\theta\var(z)
\een
Primary superfields are defined by having simple
transformation properties with respect to the superconformal
generators (\ref{LG}). With $\D$ denoting the (super)conformal
dimension, a primary superfield may be characterized by 
\bea
 \left[L_n,\Phi_\D(z,\theta)\right]&=&\left(z^{n+1}\pa_z
   +\hf(n+1)z^n\theta\pa_\theta+\D(n+1)z^n\right)\Phi_\D(z,\theta)\nn
 \left[G_r,\Phi_\D(z,\theta)\right]&=&\left(
   z^{r+1/2}\left(\pa_\theta-\theta\pa_z\right)-\D(2r+1)\theta
   z^{r-1/2}\right)\Phi_\D(z,\theta)
\label{primary}
\eea
As indicated, we are considering {\em even} primary 
superfields only.

The superconformal transformations generated
by Virasoro supergroup elements $\cg$ extend the ordinary
conformal case (\ref{GphiG}):
\ben
 \cg^{-1}\Phi_\D(z,\theta)\cg\ 
  =\ \left(D\theta'\right)^{2\D} \Phi_\D(z',\theta')
\label{GPhiG}
\een
Supergroup elements may be constructed by exponentiating
the generators of the superconformal algebra. 
Modes of the supercurrent must in that case be accompanied
by Grassmann odd parameters.

A highest-weight module of the superconformal algebra is reducible
if it contains a submodule generated from a singular
vector. The simplest non-trivial singular vector
in a highest-weight module with highest weight $\D$ 
appears at level 3/2 and is given by
\ben
 \ket{\chi;3/2}\ =\ \left((\D+\hf)G_{-3/2}-L_{-1}G_{-1/2}\right)\ket{\D}
\label{chi}
\een
provided
\ben
 12\D\ =\ (2\D+1)(3\D+c)
\label{Dc}
\een

\section{Graded stochastic evolutions and SCFT}

\subsection{General link}

Since we already have experience with stochastic calculus
of generically non-commutative objects from the Virasoro case
in section 2, for example, the inclusion of anti-commuting variables
is straightforward. An important feature is the vanishing of
$\pa_\eta^2$ ($\eta$ odd) appearing in the second-order term
in the extended or 
'graded' Ito formula. The Ito formula for commuting as well
as anti-commuting variables
is discussed in \cite{Rog} and references therein.
The techniques will be illustrated as we go.

We consider first the stochastic differential
\ben
 \cg^{-1}_td\cg_t\ =\ \al dt+\beta dB_t\ ,\ \ \ \ \ \ \cg_0=1
\label{cg}
\een
which we can think of as a random walk on the Virasoro supergroup.
$\al$ and $\beta$ are even and generically non-commutative
expressions in the generators of the superconformal algebra.
It follows from $d(\cg^{-1}_t\cg_t)=0$ that the Ito differential
of the inverse element is given by
\ben
 d(\cg^{-1}_t)\cg_t\ =\ (-\al+\beta^2)dt-\beta dB_t
\label{cg-1}
\een
The superconformal transformation generated
by $\cg_t$ acts on a primary superfield as (\ref{GPhiG}):
\ben
 \cg^{-1}_t\Phi_\D(z,\theta)\cg_t\ 
  =\ \left(D\theta'_t\right)^{2\D} \Phi_\D(z'_t,\theta'_t)
\label{GPhiGt}
\een
The superspace coordinate $(z'_t,\theta'_t)$ is now a
stochastic function of $(z,\theta)$. To relax the notation,
the subscript $t$ will often be suppressed below.

A goal is to compute the Ito differential of
both sides of (\ref{GPhiGt}) and thereby relate the
stochastic differential equations of $\cg_t$ and $(z',\theta')$.
Let's first consider the Ito differential of the left
hand side of (\ref{GPhiGt}):
\bea
 d\left(\cg^{-1}_t\Phi_\D\cg_t\right)&=&d(\cg^{-1}_t)\Phi_\D\cg_t+
  \cg^{-1}_t\Phi_\D d\cg_t+d(\cg^{-1}_t)\Phi_\D d\cg_t\nn
 &=&\left(-\left[\al-\hf\beta^2,\cg^{-1}_t\Phi_\D\cg_t\right]
  +\hf\left[\beta,\left[\beta,\cg^{-1}_t\Phi_\D\cg_t\right]\right]\right)dt\nn
   &&-\left[\beta,\cg^{-1}_t\Phi_\D \cg_t\right]dB_t
\eea
To facilitate the comparison we want 
to express the differentials in the {\em adjoint} representation
of the algebra generators only. We should thus choose
$\beta$ and $\al_0$ linear in the generators, with
$\al_0$ defined through $\al=\al_0+\hf\beta^2$.
Extending this to a $b$-dimensional Brownian motion, 
$\bar{B}_t=(B^{(1)}_t,...,B^{(b)}_t)$ with $B_0^{(i)}=0$,
we are led to consider
\ben
 \cg^{-1}_td\cg_t\ =\ \left(\al_0+\hf \sum_{i=1}^b\beta_i^2\right)dt
   +\sum_{i=1}^b\beta_i dB_t^{(i)}\ ,\ \ \ \ \ \ \cg_0=1
\label{Gal0}
\een
The associated Ito calculus treats the basic differentials according
to the rules
\ben
 (dt)^2\ =\ dtdB_t^{(i)}\ =\ 0\ ,\ \ \ \ \ 
   dB_t^{(i)}dB_t^{(j)}\ =\ \delta_{ij}dt
\een
and we find
\bea
 d\left(\cg^{-1}_t\Phi_\D \cg_t\right)
 &=&\left(-\left[\al_0,\cg^{-1}_t\Phi_\D\cg_t\right]
   +\hf\sum_{i=1}^b\left[\beta_i,\left[\beta_i,\cg^{-1}_t\Phi_\D\cg_t
  \right]\right]\right)dt\nn
 &&-\sum_{i=1}^b\left[\beta_i,\cg^{-1}_t\Phi_\D\cg_t\right]dB_t^{(i)}\nn
 &=&\left(D\theta'\right)^{2\D}\left(-\left[\al_0,\Phi_\D(z',\theta')\right]
   +\hf\sum_{i=1}^b\left[\beta_i,\left[\beta_i,\Phi_\D(z',\theta')
  \right]\right]\right)dt\nn
 &&-\left(D\theta'\right)^{2\D}
   \sum_{i=1}^b\left[\beta_i,\Phi_\D(z',\theta')\right]dB_t^{(i)}
\label{Dal0}
\eea

With the subscript $t$ suppressed, we can express the
Ito differentials of the superspace coordinate as
\bea
 dz'&=&z_0'dt+\sum_{i=1}^bz_i'dB_t^{(i)}\ ,\ \ \ \ \ \ z'|_{t=0}\ =\ z\nn
 d\theta'&=&\theta_0'dt+\sum_{i=1}^b\theta_i'dB_t^{(i)}\ ,\ \ \ \ \ \ 
  \theta'|_{t=0}\ =\ \theta\nn
 d(D\theta')&=&(D\theta_0')dt+\sum_{i=1}^b(D\theta_i')dB_t^{(i)}
\label{dzdt}
\eea
where we have added the Ito differential of $D\theta'$. The initial
conditions are required to match the initial condition on
$\cg_t$ in (\ref{Gal0}).
With (\ref{dzdt}) at hand, we compute the Ito differential
of the right hand side of (\ref{GPhiGt}):
\bea
 &&d\left(\left(D\theta'\right)^{2\D} \Phi_\D(z',\theta')\right)\nn
 &=&\left(D\theta'\right)^{2\D}\pa_{z'}\Phi_\D(z',\theta') dz'
  +\left(D\theta'\right)^{2\D}d\theta'\pa_{\theta'}\Phi_\D(z',\theta')\nn
&& +2\D\left(D\theta'\right)^{2\D-1}\Phi_\D(z',\theta') d(D\theta')
  +\hf\left(D\theta'\right)^{2\D}\pa_{z'}^2\Phi_\D(z',\theta')(dz')^2\nn
&& +\D(2\D-1)\left(D\theta'\right)^{2\D-2}\Phi_\D(z',\theta')(d(D\theta'))^2
  +\left(D\theta'\right)^{2\D}d\theta'\pa_{z'}\pa_{\theta'}
  \Phi_\D(z',\theta') dz'\nn
&& +2\D\left(D\theta'\right)^{2\D-1}\pa_{z'}\Phi_\D(z',\theta') 
     dz'd(D\theta')
  +2\D\left(D\theta'\right)^{2\D-1}d\theta'\pa_{\theta'}
  \Phi_\D(z',\theta') d(D\theta')\nn
&=&\left(D\theta'\right)^{2\D}\left\{\hf\sum_{i=1}^b(z_i')^2\pa_{z'}^2
  +\sum_{i=1}^bz_i'\theta_i'\pa_{\theta'}\pa_{z'}\right.
  +\left(z_0'+2\D\sum_{i=1}^b(D'\theta_i')z_i'
  \right)\pa_{z'}\nn
 &&+\left(\theta_0'+2\D\sum_{i=1}^b(D'\theta_i')
  \theta_i'\right)\pa_{\theta'}
 \left.+2\D\left((D'\theta_0')+(\D-\hf)\sum_{i=1}^b
  (D'\theta_i')^2\right)\right\}\Phi_\D(z',\theta')dt\nn
 && +\left(D\theta'\right)^{2\D}\sum_{i=1}^b\left\{
  z_i'\pa_{z'}+\theta_i'\pa_{\theta'}+2\D(D'\theta_i')
  \right\}\Phi_\D(z',\theta')dB_t^{(i)}
\label{dDtheta}
\eea
Note the positions of the odd differential $d\theta'$. 
We have used the transformation rule
(\ref{D'}) in the rewriting.
A comparison of the $dB_t^{(i)}$ terms in (\ref{Dal0}) and
(\ref{dDtheta}) yields
\ben
 \left[\beta_i,\Phi_\D(z',\theta')\right]\ =\ -\left(
  z_i'\pa_{z'}+\theta_i'\pa_{\theta'}+2\D(D'\theta_i')
  \right)\Phi_\D(z',\theta')
\label{dBi}
\een
With $\al_0$ and $\beta_i$ linear in the algebra generators
\bea
 \al_0&=&\sum_{n\in\Z}\left(y_{0,n}L_n+\eta_{0,n}G_{n+1/2}\right)\nn
 \beta_i&=&\sum_{n\in\Z}\left(y_{i,n}L_n+\eta_{i,n}G_{n+1/2}\right)
\label{al0beta}
\eea
we find that
\bea
 z_i'&=&-\sum_{n\in\Z}\left(y_{i,n}+\theta'\eta_{i,n}\right)(z')^{n+1}\nn
 \theta_i'&=&-\sum_{n\in\Z}\left(\hf(n+1)\theta'y_{i,n}+z'\eta_{i,n}
  \right)(z')^n
\label{ziti}
\eea
A comparison of the $dt$ terms allows us to extract information
on $\left[\al_0,\Phi_\D(z',\theta')\right]$ from which we deduce that
\bea
 z_0'&=&-\sum_{n\in\Z}\left(y_{0,n}+\theta'\eta_{0,n}\right)(z')^{n+1}
  +\hf\sum_{i=1}^b\left(z_i'\pa_{z'}+\theta_i'\pa_{\theta'}
  \right)z_i'\nn
 \theta_0'&=&-\sum_{n\in\Z}\left(\hf(n+1)\theta'y_{0,n}+z'\eta_{0,n}
  \right)(z')^n+\hf\sum_{i=1}^b\left(z_i'\pa_{z'}+\theta_i'\pa_{\theta'}
  \right)\theta_i'
\label{z0t0}
\eea
Using (\ref{ziti}), one may of course eliminate $z_i'$ and
$\theta_i'$ from these expressions, but the result is then less
compact than (\ref{z0t0}).
Only finitely many of the expansion coefficients $y$ and $\eta$
will be chosen non-vanishing rendering the sums in the solution 
(\ref{ziti}) and (\ref{z0t0}) finite.

In conclusion, the construction above establishes a general link between
a class of stochastic evolutions in superspace and SCFT:
the stochastic differentials (\ref{dzdt}) describing
the evolution of the superconformal maps are expressed
in terms of the parameters of the random walk
on the Virasoro supergroup (\ref{Gal0}), and this has
been achieved via the definition of primary
superfields in SCFT (\ref{GPhiGt}).

\subsection{Expectation values and singular vectors}

To obtain a more direct relationship 
along the lines of section 2, we should link
the representation theory of the superconformal algebra
through the construction of singular vectors to quantities
conserved in mean under the stochastic process.
{}From the supergroup differential (\ref{Gal0})
it follows that the time evolution of the
expectation value of $\cg_t\ket{\Delta}$ is given by
\ben
 \pa_t{\bf E}[\cg_t\ket{\Delta}]\ =\ {\bf E}[\cg_t\left(\al_0+
   \hf \sum_{i=1}^b\beta_i^2\right)\ket{\Delta}]
\label{EG}
\een
We should thus look for processes allowing us to put
\ben
 \left(\al_0+\hf \sum_{i=1}^b\beta_i^2\right)\ket{\D}\ \simeq \ 0
\label{0}
\een
in the representation theory. 
$\cg_t\ket{\D}$ is then a so-called martingale of the 
stochastic process $\cg_t$.

Before doing that, let us indicate how time evolutions of much
more general expectation values may be evaluated.
This extends one of the main results in \cite{BB} on ordinary
SLE. In this regard, observables of the process $\cg_t$ are
thought of as functions of $\cg_t$. We introduce the
graded vector fields
\bea
 (\nabla_nF)(\cg_t)&=&\frac{d}{du}F(\cg_te^{uL_n})|_{u=0}\nn
 (\nabla_{n+1/2}F)(\cg_t)&=&\frac{d}{d\nu}F(\cg_te^{\nu G_{n+1/2}})
   |_{\nu=0}
\label{nabla}
\eea
associated to the generators $L_n$ and $G_{n+1/2}$. Here
$\nabla_{n+1/2}$ and $\nu$ (and of course $G_{n+1/2}$)
are odd, and $F$ is a
function admitting a 'sufficiently convergent' Laurent expansion.
Referring to the notation in (\ref{al0beta}), we then find that
\ben
 \pa_t{\bf E}[F(\cg_t)]\ =\ {\bf E}[\left(\al_0(\nabla)+\hf\sum_{i=1}^b
   \beta_i^2(\nabla)\right)F(\cg_t)]
\label{nablaF}
\een
with
\bea
 \al_0(\nabla)&=&\sum_{n\in\Z}\left(y_{0,n}\nabla_n+
   \eta_{0,n}\nabla_{n+1/2}\right)\nn
 \beta_i(\nabla)&=&\sum_{n\in\Z}\left(y_{i,n}\nabla_n+
   \eta_{i,n}\nabla_{n+1/2}\right)
\label{alnabla}
\eea
This follows from (\ref{cg}), (\ref{cg-1}), 
$\al=\al_0+\hf\sum_{i=1}^b\beta_i^2$ and 
\bea
 \nabla_n(\cg^{-N}_t)&=&\frac{d}{du}
  \left(\left(e^{-uL_n}\cg_t^{-1}\right)^N\right)|_{u=0}\nn
  \nabla_{n+1/2}(\cg^{-N}_t)&=&\frac{d}{d\nu}
  \left(\left(e^{-\nu G_{n+1/2}}\cg_t^{-1}\right)^N\right)|_{\nu=0}
\eea
It is seen that (\ref{nablaF}) reduces to (\ref{EG}) as it should
when $F(\cg_t)=\cg_t\ket{\D}$.
 
We now return to (\ref{0}) and shall discuss the example
where it corresponds to the level-3/2 singular vector
(\ref{chi}). We consider two different ways of obtaining this
vector and examine the associated stochastic differential equations
in superspace. 

It turns out that a one-dimensional Brownian motion 
suffices if we choose
\ben 
 \al_0\ =\ -y\eta G_{-3/2}\ ,\ \ \ \ \ \ 
   \beta\ =\ (yL_{-1}+\eta G_{-1/2})\sqrt{\kappa}\ ,\ \ \ \ \ \ y^2=0
\label{32}
\een
The condition $y^2=0$ ensures that the term with $L_{-1}^2$
in $\beta^2$ vanishes, and is obtained by writing 
$y$ as an even monomial in Grassmann odd parameters.
$\eta$ is odd. With this choice of normalization (one can
rescale $\al_0$ by any non-vanishing real number) and by comparing
with (\ref{chi}) and (\ref{Dc}), we find
\ben
 c_\kappa\ =\ \frac{15}{2}-3\left(\kappa+\frac{1}{\kappa}\right)     
  \ ,\ \ \ \ \ \ \D_\kappa\ =\ \frac{2-\kappa}{2\kappa}
\label{cdk}
\een
This is in accordance with the ordinary labeling of reducible highest-weight
modules of the superconformal algebra with $\D=h_{1,3}$ or
$\D=h_{3,1}$. We see that the stochastic models with $\kappa$
and $1/\kappa$ may be linked to the same SCFT,
albeit via the two different primary superfields
$\Phi_{h_{1,3}}$ and $\Phi_{h_{3,1}}$.

The stochastic differential equations associated to (\ref{32}) read
\bea
 dz'&=&\frac{y\theta'\eta}{z'}dt-(y+\theta'\eta)\sqrt{\kappa}dB_t\ ,\ \ \ \ \ 
  z'|_{t=0}\ =\ z\nn
 d\theta'&=&\frac{y\eta}{z'}dt-\eta\sqrt{\kappa}dB_t\ ,\ \ \ \hspace{1.9cm}
  \theta'|_{t=0}\ =\ \theta
\label{d32}
\eea  
Unlike the similar equation in ordinary SLE (\ref{df}), 
these can actually be solved
explicitly. To see this, we introduce the functions
\bea
 w_t&=&z_t'+(y+\theta'_t\eta)\sqrt{\kappa}B_t\ ,\ \ \ \ \ \ \ w_0\ =\ z\nn
 \mu_t&=&\theta'_t+\eta\sqrt{\kappa}B_t\ ,\ \ \ \ \ \ \ \hspace{1.4cm}
     \mu_0\ =\ \theta
\eea
which are seen to satisfy the {\em ordinary} (but coupled) 
differential equations
\bea
 \pa_tw_t&=&\frac{y\mu_t\eta}{w_t}\ ,\ \ \ \ \ w_0\ =\ z\nn
 \pa_t\mu_t&=&\frac{y\eta}{w_t}\ ,\ \ \ \ \ \ \ \ \mu_0\ =\ \theta
\label{wmusimple}
\eea
{}From $\pa_t(\mu_tw_t)$ it follows that $\mu_tw_t=\theta z+y\eta t$
and in particular
\ben
 w_t\ =\ z+\frac{\theta y\eta}{z}t\ ,\ \ \ \ \ \ \ \ 
  \mu_t\ =\ \theta+\frac{y\eta}{z}t
\een
Returning to the original stochastic functions we find that they are given by
\bea
 z_t'&=&z+\frac{\theta y\eta}{z}t-(y+\theta\eta)\sqrt{\kappa}B_t\nn
 \theta'_t&=&\theta+\frac{y\eta}{z}t-\eta\sqrt{\kappa}B_t
\eea
It is easily verified that $(z,\theta)\mapsto(z'_t,\theta'_t)$ indeed
is a superconformal transformation. It is observed that the
complex part of $z$ is fixed under this map. 

A different scenario, but in some sense closer in spirit to ordinary SLE, 
emerges if we work with {\em two-dimensional} Brownian motion
(even though SLE is based on {\em one-dimensional} Brownian motion only)
and choose
\ben 
 \al_0\ =\ -y\eta G_{-3/2}\ ,\ \ \ \ \ \ 
   \beta_1\ =\ (yL_{-1}+\eta G_{-1/2})\sqrt{\kappa}\ ,\ \ \ \ \ \
  \beta_2\ =\ iyL_{-1}\sqrt{\kappa}
\label{32alt}
\een
Note that $y^2$ is {\em not} required to vanish in this case.
Nevertheless, the expression (\ref{0}) corresponds to
the same level-3/2 singular vector as in the previous case,
and we again recover the parameterization
(\ref{cdk}). The associated stochastic differentials
now read
\bea
 dz'&=&\frac{y\theta'\eta}{z'}dt-(y+\theta'\eta)\sqrt{\kappa}dB_t^{(1)}
  -iy\sqrt{\kappa}dB_t^{(2)}\ , \ \ \ \ \ \ \ \ 
  z'|_{t=0}\ =\ z\nn
 d\theta'&=&\frac{y\eta}{z'}dt-\eta\sqrt{\kappa}dB_t^{(1)}
   \ ,\ \ \ \hspace{4.3cm}\ \ \ \theta'|_{t=0}\ =\ \theta
\label{d32alt}
\eea  
These can be expressed as ordinary differential equations by introducing
\bea
 w_t&=&z_t'+(y+\theta'_t\eta)\sqrt{\kappa}B^{(1)}_t
   +iy\sqrt{\kappa}B_t^{(2)}\ ,\ \ \ \ \ \ \ \ \ \ w_0\ =\ z\nn
 \mu_t&=&\theta'_t+\eta\sqrt{\kappa}B^{(1)}_t\ ,\ \ \ \ \ \ \ \hspace{4cm}
     \mu_0\ =\ \theta
\eea
as we find that
\bea
 \pa_tw_t&=&\frac{y\mu_t\eta}{w_t-y\sqrt{\kappa}B^+_t}
  \ ,\ \ \ \ \ \ \ \ w_0\ =\ z\nn
 \pa_t\mu_t&=&\frac{y\eta}{w_t-y\sqrt{\kappa}B^+_t}
   \ ,\ \ \ \ \ \ \ \ \mu_0\ =\ \theta
\label{wmualt}
\eea
Here we have defined the complex Brownian motion
\ben
 B^+_t:=\ B^{(1)}_t+iB^{(2)}_t\ ,\ \ \ \ \ \ \ B^+_0\ =\ 0
\een
and it has been used that the complex part of $z'$ is
non-vanishing for a locally invertible superconformal map.
We note that if $y^2=0$ the set of equations (\ref{wmualt})
reduces to (\ref{wmusimple}).

Despite the resemblance to L\"owner's equation (\ref{g}), the
stochastic differential equations (\ref{wmualt}) may be solved explicitly.
To see this, we use the expansion (\ref{zt}) of $z'$ and $\theta'$
in (\ref{d32alt}) with $y$ chosen as $y=1$ for simplicity.
The four functions $g$, $\g$, $\tau$ and $s$ are then expanded
with respect to the odd parameter $\eta$ and we end up with
eight coupled differentials. They are easily solved, though, and
we find the solution to (\ref{d32alt}) to be
\bea
 z'_t&=&z-\sqrt{\kappa}B^+_t+\theta\eta\left(\int_0^t
   \frac{1}{z-\sqrt{\kappa}B^+_s}ds-\sqrt{\kappa}B_t^{(1)}
   \right)\nn
 \theta'&=&\theta+\eta\left(\int_0^t
   \frac{1}{z-\sqrt{\kappa}B^+_s}ds-\sqrt{\kappa}B_t^{(1)}
   \right)
\label{solalt}
\eea
The map $(z,\theta)\mapsto(z',\theta')$ is seen to be 
superconformal.

This solution is only well-defined when the denominator of
the integrands has a non-vanishing complex part:
$z^c\neq\sqrt{\kappa}B^+_s$, $0\leq s\leq t$, with $z^c$ denoting
the complex part of $z$.
With reference to the definition of the SLE trace in ordinary
SLE (\ref{trace}), it thus seems natural to define the 'supertrace'
$\G(t)$ to be 
\ben
 \G(t): =\ \sqrt{\kappa}B^+_t
\label{supertrace}
\een
and for $t>0$ to introduce the hulls $K_t$ as the union of
$\G[0,t]$ and the bounded connected components of $\C\setminus
\G[0,t]$. The (stochastic) superconformal maps may be interpreted
as describing the evolution of these hulls, and are taken as
maps from $(z,\theta)$ with
$z^c\in \C\setminus K_t$. Note that this also ensures
that they are locally invertible.

\section{Conclusion}

Motivated by the success of SLE in the description
of critical systems, we have extended the study in \cite{BB}
(and \cite{LR}) of the link between SLE and CFT to a link between stochastic
evolutions in superspace and SCFT. Our analysis has led to very
general results and has been illustrated by establishing a direct
relationship based on the evaluation of level-3/2 singular
vectors in highest-weight modules of the superconformal
algebra. Two different scenarios were outlined. 
A set of coupled stochastic differential
equations were solved in each of them.
In one case, the solution suggested the introduction
of a supertrace and could be interpreted as describing
the evolution of hulls in the complex plane.

Our approach may be adapted to
stochastic evolutions in 'ordinary' space and CFT.
This would extend the results of \cite{BB} and \cite{LR},
and will be discussed elsewhere.
\vskip.5cm
\noindent{\em Acknowledgements}
\vskip.1cm
\noindent The author is grateful to J. Nagi for numerous helpful discussions 
on superspace and SCFT. He also thanks C. Cummins, P. Jacob, P. Mathieu,
and Y. Saint-Aubin for comments.

\end{document}